\documentclass[sigconf]{acmart}
\usepackage[most]{tcolorbox}
\definecolor{lightgray}{gray}{0.95} 

\usepackage{graphicx}
\usepackage{soul}

\usepackage{amssymb} 
\usepackage{hhline}
\usepackage{multirow}
\usepackage[table,xcdraw]{xcolor}
\usepackage{threeparttable}
\usepackage{array}
\usepackage{adjustbox}
\usepackage{caption}
\usepackage{tikz}

\usepackage{colortbl}
\usepackage{amsfonts}

\newcolumntype{L}[1]{>{\raggedright\let\newline\\\arraybackslash\hspace{0pt}}m{#1}}
\newcolumntype{C}[1]{>{\centering\let\newline\\\arraybackslash\hspace{0pt}}m{#1}}
\newcolumntype{R}[1]{>{\raggedleft\let\newline\\\arraybackslash\hspace{0pt}}m{#1}}

\usepackage{newfloat}
\usepackage{listings}
\floatstyle{ruled}
\newfloat{listing}{tb}{lst}{}
\floatname{listing}{Listing}

\AtBeginDocument{%
  }

\setcopyright{none}
\copyrightyear{2025}
\acmYear{2025}
\acmConference[Workshop on
Recommender Systems for
Sustainable Development]{CIKM@RS4SD'25}{Nov 10--14,
  2025}{}
\acmISBN{978-1-4503-XXXX-X/2025/11}

\begin{document}

\title{Towards Trustworthy LLM-Based Recommendation \\ via Rationale Integration}

\author{Chung Park}
\orcid{1234-5678-9012}
\affiliation{%
  \institution{SK Telecom}
  \city{Seoul}
  \country{South Korea}
}
\email{cpark88kr@gmail.com}

\author{Taesan Kim}
\affiliation{%
  \institution{SK Telecom}
  \city{Seoul}
  \country{South Korea}}
\email{ktmountain@sk.com}

\author{Hyeongjun Yun}
\affiliation{%
  \institution{SK Telecom}
  \city{Seoul}
  \country{South Korea}}
\email{hjyoon@sk.com}

\author{Dongjoon Hong}
\affiliation{%
  \institution{SK Telecom}
  \city{Seoul}
  \country{South Korea}}
\email{dongjoon.hong@sk.com}

\author{Junui Hong}
\affiliation{%
  \institution{SK Telecom}
  \city{Seoul}
  \country{South Korea}}
\email{skt.juhong@sk.com}

\author{Kijung Park}
\affiliation{%
  \institution{SK Telecom}
  \city{Seoul}
  \country{South Korea}}
\email{kijung.park@sk.com}

\author{MinCheol Cho}
\affiliation{%
  \institution{SK Telecom}
  \city{Seoul}
  \country{South Korea}}
\email{skt.mccho@sk.com}

\author{Mira Myong}
\affiliation{%
  \institution{SK Telecom}
  \city{Seoul}
  \country{South Korea}}
\email{mira.myong@sk.com}

\author{Jihoon Oh}
\affiliation{%
  \institution{SK Telecom}
  \city{Seoul}
  \country{South Korea}}
\email{skt.jihoon@sk.com}

\author{Min sung Choi}
\affiliation{%
  \institution{SK Telecom}
  \city{Seoul}
  \country{South Korea}}
\email{ms.choi@sk.com}


\renewcommand{\shortauthors}{Chung Park et al.}

\begin{abstract}
Traditional recommender systems (RS) have been primarily optimized for accuracy and short-term engagement, often overlooking transparency and trustworthiness.
Recently, platforms such as Amazon and Instagram have begun providing recommendation rationales to users, acknowledging their critical role in fostering trust and enhancing engagement; however, most existing systems still treat them as post-hoc artifacts.
We propose an LLM-based recommender (LLM-Rec) that not only predicts items but also generates logically grounded rationales. Our approach leverages a self-annotated rationale dataset and instruction tuning in a rationale-first format, where the model generates an explanation before outputting the recommended item.
By adopting this strategy and representing rationales in a chain-of-thought (CoT) style, LLM-Rec strengthens both interpretability and recommendation performance.
Experiments on the Fashion and Scientific domains of the Amazon Review dataset demonstrate significant improvements over well-established baselines. 
To encourage reproducibility and future research, we publicly release a rationale-augmented recommendation dataset containing user histories, rationales, and recommended items \footnote{{\color[HTML]{0037D7}\url{https://drive.google.com/drive/u/1/folders/1kIQQQbUZKStKWb_AwFPijaSfFKVsS_H5}}}.
\end{abstract}





\maketitle

\section{Introduction}
Recommender systems play a critical role in shaping consumer choices across shopping, media, and digital services~\cite{park2024pacer}. 
Traditional approaches such as collaborative filtering have achieved strong performance, but they are predominantly optimized for short-term accuracy and engagement, with limited consideration for transparency, trustworthiness, or broader societal objectives.
Recently, platforms such as \textit{Amazon} and \textit{Instagram} have introduced recommendation rationales—concise natural language explanations designed to increase transparency by clarifying why items are recommended.
These rationales improve transparency and user trust, yet most existing systems treat them as post-hoc artifacts rather than integrating them into the learning and reasoning pipeline~\cite{lei2024recexplainer}.

Advances in Large Language Models (LLMs) provide a timely opportunity to unify recommendation and rationale generation within a single framework. 
In this work, we propose a rationale-aware LLM-based recommender (LLM-Rec) trained via instruction tuning on self-annotated rationale datasets, where each instance is structured to first generate a rationale and subsequently output the recommended item. 
By adopting this rationale-first decoding strategy, the model leverages its reasoning capacity not only to deliver coherent explanations but also to enhance predictive performance. 
\textbf{Furthermore, we show that representing rationales in a chain-of-thought (CoT) style yields benefits from both train-time scaling and test-time scaling perspectives, offering new insights into how rationale generation itself contributes to recommendation performance.}
We validate our approach on two contrasting domains of the Amazon Review dataset—Fashion and Scientific. 
Experiments demonstrate that rationale-aware learning and inference yield significant improvements, confirming that rationales serve not only as a means of interpretability and business utility but also as a direct driver of recommendation performance. 

\vspace{-0.2cm}
\section{Methodology}
\subsection{Rationale-Enriched Data Construction}
We adopt a sequential recommendation setting where a user’s historical interactions $S_u$ and ground-truth next item $i^+$ form the base dataset. Using a state-of-the-art LLM (e.g., \texttt{gpt-4o}), we generate post-hoc rationales $r_{i^+}$ explaining why the next item is coherent with $S_u$. 
The model is prompted to flag incoherent explanations, which are filtered out. 
Final training tuples are of the form $(S_u, i^+, r_{i^+})$.  

\subsection{Rationale-Aware Instruction Tuning}
The rationale-augmented dataset is used to instruction-tune an LLM recommender. During training and inference, the model is prompted to follow a rationale-first decoding process: infer user preferences, generate a rationale, and then recommend an item. 
To enable seamless integration into real-world recommender system pipelines, we trained the model to follow a structured output format: the rationale is enclosed within \texttt{<think>} \texttt{</think>} tags, and the recommended item within \texttt{<item>} \texttt{</item>} tags.
This chain-of-thought structure injects reasoning into the recommendation process.
In subsequent stages, additional alignment training (e.g., Group Relative Policy Optimization) may be applied to further refine the model.
An example of the prompt for this task is as follows.

\vspace{-0.2cm}
\begin{tcolorbox}[colback=white, boxrule=0.8pt, arc=5pt,
  top=0.02pt, bottom=0.02pt, left=0.02pt, right=0.02pt,  
  boxsep=0.5pt,  
  enhanced]
  \begin{tcolorbox}[colback=lightgray, colframe=black, boxrule=0.6pt,
    arc=3pt, boxsep=3pt,  
    left=1pt, right=1pt, top=1pt, bottom=1pt]  
  \tiny 
\setlength{\fboxsep}{0.8pt}\colorbox[HTML]{f4d03f}{\textbf{\#Instruction}} \\
Based on \textbf{{[}Purchase History{]}}, use your logical reasoning process to identify the most suitable item for this customer from the \textbf{{[}Candidate List{]}}.
Then, include your reasoning inside the \textless{}\texttt{think}\textgreater \textless{}\texttt{/think}\textgreater tag and the recommended item inside the \textless{}\texttt{item}\textgreater \textless{}\texttt{/item}\textgreater tag.\\
\setlength{\fboxsep}{0.8pt}\colorbox[HTML]{f4d03f}{\textbf{\#Input}} \\
\textbf{{[}Purchase History{]}} (1)Title: Classic fuzzy ribbed knit beanie hat  (2)Title: Hotstyle bestie mini backpack purse,...(truncated)  \setlength{\fboxsep}{0.8pt}\colorbox[HTML]{9FE2BF}{$\leftarrow S_{u}$}\\
\textbf{{[}Candidate List{]}}: (omitted for brevity)\\
\setlength{\fboxsep}{0.8pt}\colorbox[HTML]{f4d03f}{\textbf{\#Output}} \\
\textless{}\texttt{think}\textgreater Based on the customer’s history of purchasing fashion accessories like the classic fuzzy ribbed knit beanie, the system recommends...(truncated) \textless{}\texttt{/think}\textgreater  \setlength{\fboxsep}{0.5pt}\colorbox[HTML]{9FE2BF}{$\leftarrow r_{i^{+}}$}\\
\textless{}\texttt{item}\textgreater Oalka women's joggers high waist yoga pockets sweatpants sport workout pants \textless{}\texttt{/item}\textgreater  \setlength{\fboxsep}{0.8pt}\colorbox[HTML]{9FE2BF}{$\leftarrow i^{+}$}
  \end{tcolorbox}
\end{tcolorbox}

\vspace{-0.4cm}
\section{Datasets}
We select two representative domains from the Amazon Review Dataset 2023~\cite{hou2024bridging}.
The \textbf{Grocery and Gourmet Food} domain contains 11,334 users and 6,149 items, with user histories averaging 10.76 interactions.
The \textbf{Office Products} consists of 11,810 users and 8,965 items, where user histories average 6.94 interactions.

\vspace{-0.2cm}
\begin{table}[h] \small 
\centering
\caption{Model comparison, highlighting top two methods.
}
\setlength{\tabcolsep}{3.4pt}
\renewcommand{\arraystretch}{0.8}
\label{tab:result_1}
\begin{tabular}{cc|ccc|ccc}
\toprule
\multicolumn{2}{c|}{\textbf{Domain}}
  & \multicolumn{3}{c|}{\textbf{Grocery}}
  & \multicolumn{3}{c}{\textbf{Office}} \\ \midrule
\multicolumn{1}{c|}{\textbf{Types}} & \textbf{Models}
  & HR@1 & HR@5 & N@5
  & HR@1 & HR@5 & N@5 \\ \midrule
\multicolumn{1}{c|}{} & SASRec~\cite{kang2018self}
  & 3.02 & 15.11 & 8.91
  & 2.92 & 13.76 & 8.19 \\
\multicolumn{1}{c|}{} & BERT4Rec~\cite{sun2019bert4rec}
  & 2.98 & 14.68 & 8.69
  & 3.07 & 13.21 & 8.04 \\
\multicolumn{1}{c|}{} & S3Rec~\cite{zhou2020s3}
  & 2.80 & 13.85 & 8.12
  & 2.77 & 12.63 & 7.60 \\
\multicolumn{1}{c|}{} & NextItNet~\cite{yuan2019simple}
  & 2.93 & 15.04 & 8.82
  & \underline{3.15} & 13.45 & 8.22 \\
\multicolumn{1}{c|}{\multirow{-5}{*}{\begin{tabular}[c]{@{}c@{}}CF\\ Rec\end{tabular}}}
 & LightSANs~\cite{fan2021lighter}
  & 3.08 & 14.61 & 8.75
  & 3.03 & 13.53 & 8.15 \\ \hline
\multicolumn{1}{c|}{} & SAID~\cite{hu2024enhancing}
  & 0.94 & 3.97 & 2.35
  & 0.96 & 3.64 & 2.19 \\
\multicolumn{1}{c|}{} & Rec-R1~\cite{lin2025rec}
  & 5.65 & 17.95 & 11.94
  & 2.59 & 7.87 & 5.45 \\
\multicolumn{1}{c|}{} & DPO~\cite{rafailov2023direct}
  & 3.84 & 15.19 & 9.89
  & 2.96 & 11.58 & 7.33 \\
\multicolumn{1}{c|}{} & S-DPO~\cite{chen2024softmax}
  & \underline{6.57} & \underline{18.47} & \textbf{12.91}
  & 2.59 & \underline{14.54} & \underline{8.47} \\
\multicolumn{1}{c|}{\multirow{-6}{*}{\begin{tabular}[c]{@{}c@{}}LLM\\ Rec\end{tabular}}}
 & \cellcolor[HTML]{EFEFEF}\textbf{Ours}
  & \cellcolor[HTML]{EFEFEF}\textbf{6.83} & \cellcolor[HTML]{EFEFEF}\textbf{18.53} & \cellcolor[HTML]{EFEFEF}\underline{12.58}
  & \cellcolor[HTML]{EFEFEF}\textbf{5.14} & \cellcolor[HTML]{EFEFEF}\textbf{16.11} & \cellcolor[HTML]{EFEFEF}\textbf{10.78} \\

\bottomrule
\end{tabular}
\end{table}

\vspace{-0.4cm}
\section{Discussion}
\subsection{Main Results}
\noindent\textbf{$\blacksquare$ Experiment Setting}
We compare our model against:  
(1) \textbf{CF-Rec}: collaborative filtering baselines;  
(2) \textbf{LLM-Rec}: instruction-tuned LLM recommenders without rationale supervision.  
We adopt \texttt{Gemma-3-1/4B-it}~\cite{team2025gemma} as the backbone, balancing performance and deployment feasibility. 
Evaluation follows leave-one-out: the last item is reserved for testing, the second-to-last for validation, and the rest for training. Candidate sets are formed with ground-truth plus negatives. Metrics include Hit Ratio (HR) and Normalized Discounted Cumulative Gain (NDCG).  

\noindent\textbf{$\blacksquare$ Recommendation Performance}
As shown in Table~\ref{tab:result_1}, our model, trained with self-constructed rationales, outperforms existing baselines.
Across two domains, it achieves superior results over CF- and LLM-Rec models in terms of HR@1, HR@5, and NDCG@5.

\begin{figure}[h!]
    \begin{center}
    \includegraphics[width=0.99\linewidth]{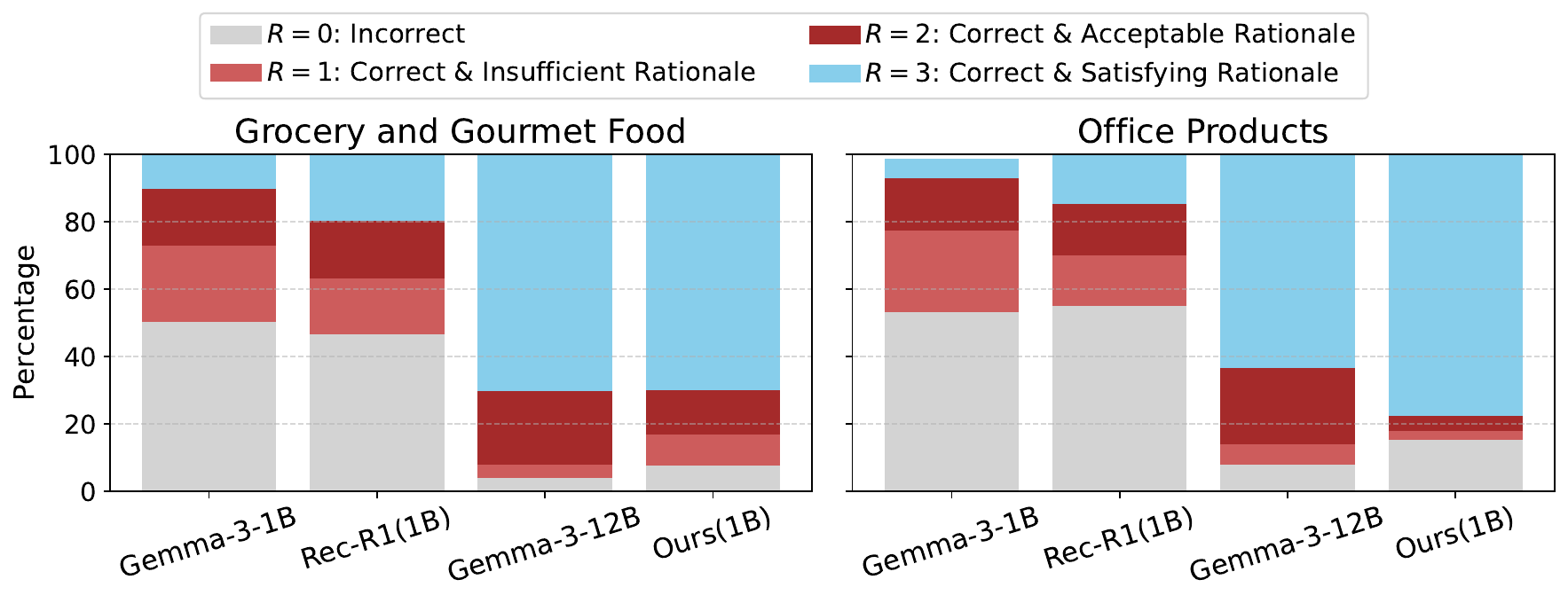}
    \end{center}
    \caption{GPT-4o Evaluation of Rationale Quality
}
    \label{fig:ablation_rationales_quality}
    \end{figure}

\begin{figure}[h!]
    \begin{center}
    \includegraphics[width=0.99\linewidth]{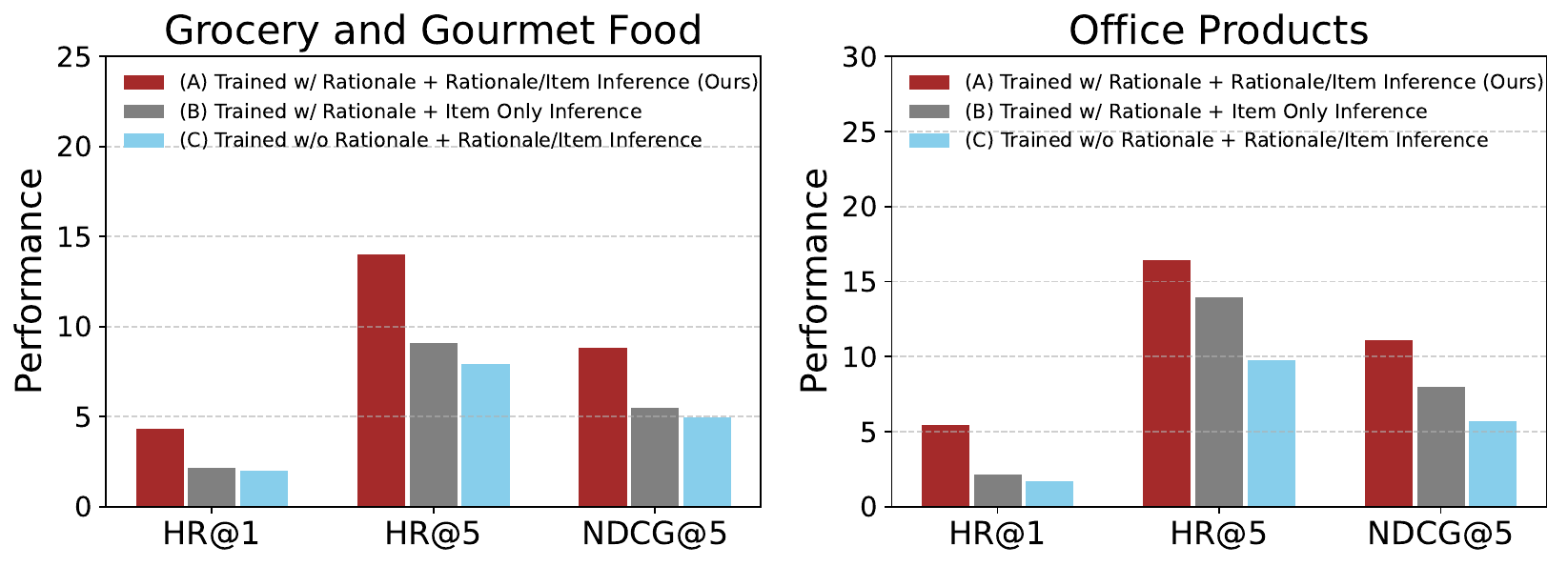}
    \end{center}
    \caption{Comparison of three training–inference variants with/without rationale
}
    \label{fig:rationale_analysis}
    \end{figure}

\noindent\textbf{$\blacksquare$ Quality of Rationales}
We evaluated rationale quality across \texttt{Gemma-3-1B}, \texttt{Rec-R1(1B)}, \texttt{Gemma-3-12B}, and ours using a unified same prompt (Fig.~\ref{fig:rationale_analysis}). 
From two Amazon domains, we sampled 600 instances (300 per domain). 
Evaluation followed a two-step protocol: (1) correctness of recommendation (item in vocabulary set) and (2) rationale quality. ChatGPT, following \citet{wang2022self}, scored outputs on a four-level scale (0–3). As shown in Fig.\ref{fig:ablation_rationales_quality}, our model, despite its relatively small 1B size, achieved the highest proportion of top-quality rationales (score 3) at 75.71\%, surpassing \texttt{Gemma-3-12B} by ~4\%p.

\vspace{-0.2cm}

\vspace{-0.45cm}
\subsection{Effect of Incorporating Rationales}
We compare three variants:  
\noindent(A) Rationale+Item (Inference), Trained w/ Rationale
\noindent(B) Item-Only (Inference), Trained w/ Rationale
\noindent(C) Rationale+Item (Inference), Trained w/o Rationale.

\noindent\textbf{$\blacksquare$ Test-Time Scaling.} The model variant (A) outperforms (B) across both domains, showing that generating rationales at inference improves HR@1 by encouraging reasoning before prediction.  

\noindent\textbf{$\blacksquare$ Training Supervision.} The model variant (A) outperforms (C), demonstrating that rationale supervision during training provides stronger signals, improving generalization.  

Our findings highlight that rationales are not merely interpretability artifacts---they actively improve both model performance and user experience. From a responsible AI perspective, rationales foster transparency, build trust, and encourage reflective choices.

\vspace{-0.2cm}

\vspace{-0.3cm}
\subsection{Online A/B Test}
We deployed the model in production and conducted an A/B test over eight weeks (July–August 2025).
Customers received top-1 recommendations with rationales, while baselines offered no explanations. 
Our model achieved 1.48× and 1.61× higher CTR than the human curation and traditional ML baselines, respectively, while maintaining practical inference latency.

\vspace{-0.2cm}
\section{Conclusion}  
We propose a rationale-aware LLM recommender that improves accuracy across two Amazon domains, while also enhancing trust and fostering reflective user behavior.

\begin{acks}
We would like to thank SK Telecom for providing access to the GPU cluster used in this research.
\end{acks}

\bibliographystyle{ACM-Reference-Format}
\bibliography{sample-base}

@String{Computing = "Computing" }

@String{Chelsea = "Chelsea" }

@ArtifactSoftware{R,
    title = {R: A Language and Environment for Statistical Computing},
    author = {{R Core Team}},
    organization = {R Foundation for Statistical Computing},
    address = {Vienna, Austria},
    year = {2019},
    url = {https://www.R-project.org/},
}

@inproceedings{park2024pacer,
  title={Pacer and Runner: Cooperative Learning Framework between Single-and Cross-Domain Sequential Recommendation},
  author={Park, Chung and Kim, Taesan and Yoon, Hyungjun and Hong, Junui and Yu, Yelim and Cho, Mincheol and Choi, Minsung and Choo, Jaegul},
  booktitle={Proceedings of the 47th International ACM SIGIR Conference on Research and Development in Information Retrieval},
  pages={2071--2080},
  year={2024}
}

@inproceedings{fan2021lighter,
  title={Lighter and better: low-rank decomposed self-attention networks for next-item recommendation},
  author={Fan, Xinyan and Liu, Zheng and Lian, Jianxun and Zhao, Wayne Xin and Xie, Xing and Wen, Ji-Rong},
  booktitle={Proceedings of the 44th international ACM SIGIR conference on research and development in information retrieval},
  pages={1733--1737},
  year={2021}
}

@inproceedings{yuan2019simple,
  title={A simple convolutional generative network for next item recommendation},
  author={Yuan, Fajie and Karatzoglou, Alexandros and Arapakis, Ioannis and Jose, Joemon M and He, Xiangnan},
  booktitle={Proceedings of the twelfth ACM international conference on web search and data mining},
  pages={582--590},
  year={2019}
}

@inproceedings{kang2018self,
  title={Self-attentive sequential recommendation},
  author={Kang, Wang-Cheng and McAuley, Julian},
  booktitle={2018 IEEE international conference on data mining (ICDM)},
  pages={197--206},
  year={2018},
  organization={IEEE}
}

@inproceedings{sun2019bert4rec,
  title={BERT4Rec: Sequential recommendation with bidirectional encoder representations from transformer},
  author={Sun, Fei and Liu, Jun and Wu, Jian and Pei, Changhua and Lin, Xiao and Ou, Wenwu and Jiang, Peng},
  booktitle={Proceedings of the 28th ACM international conference on information and knowledge management},
  pages={1441--1450},
  year={2019}
}

@inproceedings{zhou2020s3,
  title={S3-rec: Self-supervised learning for sequential recommendation with mutual information maximization},
  author={Zhou, Kun and Wang, Hui and Zhao, Wayne Xin and Zhu, Yutao and Wang, Sirui and Zhang, Fuzheng and Wang, Zhongyuan and Wen, Ji-Rong},
  booktitle={Proceedings of the 29th ACM international conference on information \& knowledge management},
  pages={1893--1902},
  year={2020}
}

@inproceedings{hu2024enhancing,
  title={Enhancing sequential recommendation via llm-based semantic embedding learning},
  author={Hu, Jun and Xia, Wenwen and Zhang, Xiaolu and Fu, Chilin and Wu, Weichang and Huan, Zhaoxin and Li, Ang and Tang, Zuoli and Zhou, Jun},
  booktitle={Companion Proceedings of the ACM on Web Conference 2024},
  pages={103--111},
  year={2024}
}

@article{lin2025rec,
  title={Rec-r1: Bridging generative large language models and user-centric recommendation systems via reinforcement learning},
  author={Lin, Jiacheng and Wang, Tian and Qian, Kun},
  journal={arXiv preprint arXiv:2503.24289},
  year={2025}
}

@article{hou2024bridging,
  title={Bridging Language and Items for Retrieval and Recommendation},
  author={Hou, Yupeng and Li, Jiacheng and He, Zhankui and Yan, An and Chen, Xiusi and McAuley, Julian},
  journal={arXiv preprint arXiv:2403.03952},
  year={2024}
}

@inproceedings{lei2024recexplainer,
  title={Recexplainer: Aligning large language models for explaining recommendation models},
  author={Lei, Yuxuan and Lian, Jianxun and Yao, Jing and Huang, Xu and Lian, Defu and Xie, Xing},
  booktitle={Proceedings of the 30th ACM SIGKDD Conference on Knowledge Discovery and Data Mining},
  pages={1530--1541},
  year={2024}
}

@article{team2025gemma,
  title={Gemma 3 technical report},
  author={Team, Gemma and Kamath, Aishwarya and Ferret, Johan and Pathak, Shreya and Vieillard, Nino and Merhej, Ramona and Perrin, Sarah and Matejovicova, Tatiana and Ram{\'e}, Alexandre and Rivi{\`e}re, Morgane and others},
  journal={arXiv preprint arXiv:2503.19786},
  year={2025}
}

@article{wang2022self,
  title={Self-instruct: Aligning language models with self-generated instructions},
  author={Wang, Yizhong and Kordi, Yeganeh and Mishra, Swaroop and Liu, Alisa and Smith, Noah A and Khashabi, Daniel and Hajishirzi, Hannaneh},
  journal={arXiv preprint arXiv:2212.10560},
  year={2022}
}

@article{chen2024softmax,
  title={On softmax direct preference optimization for recommendation},
  author={Chen, Yuxin and Tan, Junfei and Zhang, An and Yang, Zhengyi and Sheng, Leheng and Zhang, Enzhi and Wang, Xiang and Chua, Tat-Seng},
  journal={Advances in Neural Information Processing Systems},
  volume={37},
  pages={27463--27489},
  year={2024}
}

@article{rafailov2023direct,
  title={Direct preference optimization: Your language model is secretly a reward model},
  author={Rafailov, Rafael and Sharma, Archit and Mitchell, Eric and Manning, Christopher D and Ermon, Stefano and Finn, Chelsea},
  journal={Advances in neural information processing systems},
  volume={36},
  pages={53728--53741},
  year={2023}
}

\end{document}